\documentclass[10pt,journal,letterpaper,twocolumn,twoside]{IEEEtran}   

\usepackage{amsmath,amsfonts,bm,amssymb,psfrag,ifthen,color,subfigure}
\usepackage{gensymb}
\usepackage{mathrsfs}
\usepackage[justification=centering,font=footnotesize]{caption}
\usepackage{todonotes}
\usepackage{caption}

\usepackage{graphicx}
\usepackage{subfigure}
\usepackage[dvips]{epsfig}
\usepackage[dvips]{graphicx}
\usepackage{amsmath,amsfonts,bm,amssymb,psfrag,ifthen,color,subfigure}
\usepackage[ruled,linesnumbered]{algorithm2e}
\usepackage{ifthen}
\usepackage{setspace}
\usepackage{cite}
\usepackage{url}
\usepackage{longtable}
\usepackage{stfloats}  
\usepackage{graphics,booktabs,color,subfigure}
\usepackage{float}
\usepackage{diagbox}

\usepackage[dvips]{epsfig}
\usepackage[dvips]{graphicx}
\usepackage{amsmath,amsfonts,bm,amssymb,psfrag,ifthen,color,subfigure,amsthm}
\usepackage{setspace}
\usepackage{cite}
\usepackage{url}
\usepackage{longtable}
\usepackage{stfloats}  
\usepackage{graphics,booktabs,color,subfigure}
\usepackage{float}
\usepackage{diagbox}
\usepackage{multirow}

\allowdisplaybreaks[4]

\usepackage{mathrsfs}

\usepackage{graphicx}
\usepackage{makecell}
\usepackage{epstopdf}

\usepackage{graphicx}

\begin{document}
	\title{Task-Oriented Semantic Communication with Importance-Aware Rate Control}

	\author{Zhiye Sun, Shuai Ma and Shiyin Li
	}
	\maketitle
	\begin{abstract}
		Semantic communication is recognized for its high compression efficiency and robust resistance to noise. However, utilizing a fixed transmission rate in environments with dynamic signal-to-noise ratios (SNR) often results in inefficient use of communication resources. 
        To address this challenge, this letter proposes an importance-aware rate control semantic communication (IRCSC) scheme, which dynamically adjusts transmission rates in response to both channel conditions and semantic importance. 
        The scheme employs a contribution-based importance analyzer to rank semantic importance. Additionaly, a novel metric,  the semantic transmission integrity index (STII), is proposed to quantify the amount of correctly transmitted information and to correlate it with inference performance.
        Simulations indicate that, with low computational complexity, IRCSC guarantees a controllable trade-off between performance and rate, delivering higher compression efficiency and improved task performance in high-SNR scenarios.
	\end{abstract}


	\IEEEpeerreviewmaketitle
	
\section{ Introduction }
    With the rapid development of various intelligent devices, such as large-scale Internet of Things (IoT), multi-sensory extended reality (XR), autonomous driving, drones, and holographic communication \cite{akyildiz2022metaverse}, the 5G communication network is facing numerous bottlenecks. These include channel capacities approaching Shannon's limit, source coding efficiencies nearing the Shannon entropy limit \cite{strinati20196g}, massive power consumption in communication systems, and the growing scarcity of high-quality spectrum resources. Semantic communication, as an intelligent communication method whose primary objective is to convey the intended meaning of a message and to facilitate its comprehension prior to its transmission, has the capacity to enhance the transmission efficiency of communication systems and has the potential to address the technological challenges of 5G wireless networks \cite{dong2022semantic}.
  
    Recent advancements in semantic communication have introduced various methods to enhance downstream inference tasks. Information bottleneck (IB) techniques compress semantic features by balancing mutual information between input-output and intermediate representations \cite{shao2021learning}. A triplet-based semantic text representation has been proposed to improve performance in question answering (QA) and sentiment analysis \cite{liu2024explainable}. Additionally, a deep joint source-channel coding approach has been developed for wireless image retrieval \cite{jankowski2020deep}. In speech recognition, the DeepSC-SR framework utilizes convolutional neural networks (CNN) and recurrent neural networks (RNN) to boost performance \cite{weng2023deep}. However, a common limitation is the reliance on a fixed transmission rate, which may not adapt well to dynamic channel conditions.

    To adapt to variations in source data and environmental distributions, 
     the development of a fidelity-adjustable semantic transmission framework (FAST) has addressed the concept of on-demand semantic communication in heterogeneous networks \cite{li2023fast}.
      A scheme that dynamically adjusts quantization bits based on semantic importance using reinforcement learning (RL) has been proposed \cite{liu2024ofdm}. Additionally, a method combining RL with a Hungarian algorithm to allocate transmission rates and channel resources is presented in \cite{wang2024feature}. The study in \cite{wang2025channel} introduces a neural network (NN) structure to compute transmission rates based on the signal-to-noise ratio (SNR).   
       However, a significant limitation persists, existing studies often introduce complex mechanisms that incur additional computational overhead. Moreover, current research predominantly focuses on the optimal trade-off between transmission rate and performance, while neglecting to explore the explicit relationship between semantic importance and inference performance.

    The primary contributions of our letter are as follows:

    \begin{itemize}

    \item
    We propose a contribution-based importance analyzer, which enables efficient importance evaluation without incurring excessive computational overhead.

    \item
    We introduce the semantic transmission integrity index (STII) as a novel evaluation metric to measure the amount of effective information successfully transmitted in semantic communication systems. The STII integrates two critical factors: semantic importance and bit error rate (BER). Furthermore, we establish a mapping function between the STII and the inference performance of downstream tasks.
    
    \item
    Adaptive rate control is achieved in response to varying channel conditions. The proposed algorithm is characterized by low computational complexity and enables flexible control of the rate-performance trade-off according to performance requirement.

    \end{itemize}
    
    \begin{figure*}[t]
        \centering
        \includegraphics[width=\linewidth]{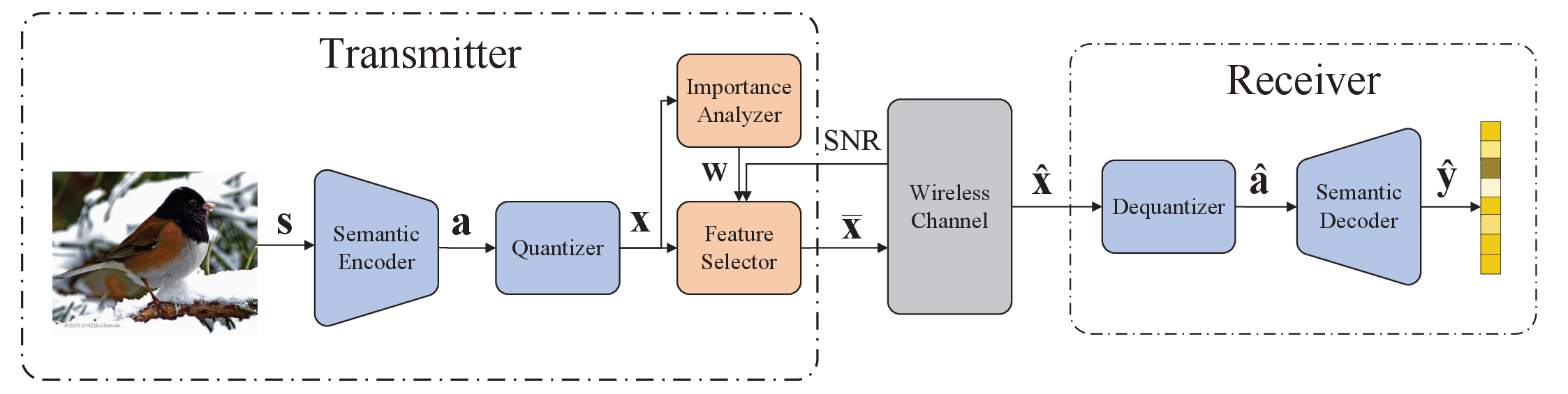}
        \caption{Structure of the proposed IRCSC system}
        \label{fig:structure}
    \end{figure*}
    
\section{ System Model }

    In this letter, we propose an end-to-end semantic communication system integrating semantic feature extraction, compression and transmission, as shown in Fig.~\ref{fig:structure}. The system consists of three parts: the transmitter, the wireless channel, and the receiver. The system is parameterized by neural networks and trained end-to-end using a data-driven approach with a back-propagation optimization strategy.

    The input image, denoted as $ \mathbf{s} \in \mathbb{R} ^ { 3 \times H \times W} $, represents an RGB image with length and width of $H$ and $W$, respectively. The transmitter first extracts the high-dimensional semantic features $ \mathbf{a} $ through a semantic encoder $F_\theta \left(  \cdot  \right) $ as
    \begin{align}
    \mathbf {a} = F_ \theta \left( \mathbf {s} \right).
    \end{align}

    The continuous semantic feature representation $\mathbf{a}$ requires analog modulation or full-resolution constellation for transmission, which imposes significant overhead on resource-constrained RF systems. Therefore, $\mathbf{a}$ is quantized as    
    \begin{align}
        \mathbf{x} = { \Psi }\left( {{\bf{a}}} \right)\label{quantizer},
    \end{align}
where $ \Psi \left(  \cdot  \right) $ is a quantizer parameterized by neural networks. The quantized semantic feature $ \mathbf{x} \in \mathbb{R} ^ { L \times C} $, where $L$ is the length of the features and $C$ is the size of the feature channel dimension, is then fed into the importance analyzer to obtain an importance weight vector $\mathbf{w} \in \mathbb{R} ^ {C} $ indicating the importance distribution of $ \mathbf {x} $ in the feature channel dimension. After sorting $\mathbf{w}$ to obtain $\overline{\mathbf{w}}$, we select the top $M$ features using a binary mask $\mathbf{m} \in \{0,1\}^C$, where $m_i=1$ for the $M$ most important features and 0 otherwise. The filtered semantic feature $\overline{\mathbf{x}} \in \mathbb{R}^{L \times M}, M \leq C$ is obtained via 
\begin{align}
\overline{\mathbf{x}} = \mathbf{x} \odot \mathbf{m}, 
\label{eq:filter}
\end{align}
the detailed process will be introduced in Section III.
Then $\overline{\mathbf{x}}$ is subsequently transmitted through a wireless channel:

    \begin{align}
        \mathbf{ \hat  { \mathbf{ x } } }  = h  \mathbf{ \overline { x } } + \mathbf {n},
    \end{align}
    where $ h $ is the channel gain between the transmitter and receiver, and $ \mathbf { n } $ is the additive white Gaussian noise (AWGN) with variance $ \sigma_ n $.

    At the receiver, the signal undergoes equalization and dequantization before being fed into the semantic decoder for downstream task inference, yielding the predicted result $ \hat { \mathbf {y} } $. This process can be expressed as

   \begin{align}
        \mathbf{ \hat  { \mathbf{ y } } }  = F ^ {-1} _ \theta \left( \Psi ^ {-1} \left(  \frac { \hat  { \mathbf{ x } } } { h } \right) \right),\label{eq:decode}
    \end{align}
    where $ F ^ { -1 } _\theta \left(  \cdot  \right) $ and $ \Psi ^ { -1 } \left(  \cdot  \right) $ represent the semantic decoder and dequantizer respectively.

\section{ Proposed IRCSC scheme }

   This section introduces an adaptive rate control scheme based on semantic importance with two key modules: an importance analyzer that quantifies feature channel distributions and a feature selector that determines transmission data volume using STII. Collectively, these modules enable robust semantic communication with high compression under varying SNR conditions.

    \subsection{ Importance analyzer }
        The quantizer outputs a semantic feature vector
        $ \mathbf{x} = [ \mathbf{x}^1, \mathbf{x}^2, ..., \mathbf{x}^C ] $, where different feature channels carry semantic information of varying importance. Therefore, we design an importance analyzer to compute the contribution of each feature channel. The importance analyzer leverages shared model weights with both the dequantizer and the semantic decoder. Given the input semantic feature $ \mathbf{x} $, the analyzer computes the classification output $ \mathbf{ \hat{y} } $, then importance weight of the semantic feature along the feature channel dimension $w_k$ can be mathematically formulated as
        
        \begin{align}
        w_k = \sum_{i=1}^L \frac{ \partial{ \hat{ y} ^*} }{ \partial{ \mathbf{ x } ^k _i } }, k=1,2,...,C,\label{eq:importance}
        \end{align}
        where $ \hat{ y} ^* $ denotes the predicted probability corresponding to the ground truth label in $ \mathbf{ \hat{y} } $. Specifically, in a multi-class classification neural network, $ \hat{ y} ^* $ represents the model's predicted probability for the correct class. This importance weight quantifies the relative contribution of individual feature channels to the downstream task results. The underlying principle relies on computing a local linear approximation of the neural network's behavior around a specific operating point \cite{selvaraju2017grad}, which can be expressed as
        \begin{align}
       \mathbf{ \hat  { \mathbf{ y } } }  \approx F ^ {-1} _ \theta \left( \Psi ^ {-1} \left(\mathbf{ x }  \right) \right) - w_k F ^ {-1} _ \theta \left( \Psi ^ {-1} \left(\mathbf{ x } | \mathbf{ x }^k = 0  \right) \right).
       \end{align}

        Subsequently, we employ merge sort to obtain the sorted importance weights $\mathbf{\overline{w}}$. 
        The computational efficiency of this module is noteworthy, as it leverages the automatic computation of derivatives during the backpropagation process. The architecture encompasses fully connected layers in both the quantizer and semantic decoder. Given that the quantizer has input and output dimensions of $P$ and $Q$ respectively, and the model predicts $K$ classes, the incorporation of the importance analyzer introduces a computational complexity of $O \left( 2PQ + PK\right)$. When combined with the merge sort algorithm's complexity of $O \left(C \log C \right)$, the overall computational complexity is thereby $O \left( 2PQ + PK + C \log C \right)$.

    \subsection{ Feature selector }
        In semantic communication systems, bit errors are generally tolerable, allowing the transmission rate to be appropriately reduced to conserve power consumption while maintaining the performance of downstream tasks. 
        
        In conventional communication systems, BER serves as the primary metric for assessing communication quality, given their focus on bit-level data transmission. However, this metric proves to be inadequate for semantic communication systems, where semantic features exhibit heterogeneous importance distributions that demand more nuanced evaluation. To overcome this limitation, we propose a novel metric called STII based on BER and semantic importance, defined as

        \begin{equation}
            \eta = \frac{{\sum\limits_{i = 1}^M {{\overline w_i}} (1 - {P_e(\rm{SNR})}) + P^0({\sum\limits_{i = 1}^C {{\overline w_i}} } - \sum\limits_{i = 1}^M {{\overline w_i}} )}}{{\sum\limits_{i = 1}^C {{\overline w_i}} }},
            \label{eq:STII}
        \end{equation}
       where $ \{ {\overline w_i} \} _ {i=1} ^ C$ represents the $i$-th element of the sorted importance distributions $\mathbf{\overline{w}}$, and $M$ is the number of selected feature channels. $P_e(\rm{SNR})$ denotes the BER. $P^0$ represents the probability of inherent predictability, which equals 0.5 under binary phase shift keying (BPSK) modulation. For semantic features transmitted through a Rayleigh channel, the BER can be expressed as

        \begin{equation}
            {P_e(\rm{SNR})} = \frac{1}{2}(1 - \sqrt {\frac{\rm{SNR}}{{1 + \rm{SNR}}}} ).
        \end{equation}

        The STII metric consists of two components. The first component, ${\sum\limits_{i = 1}^M {\overline w_i}} (1 - {P_e(\rm{SNR})})$, measures the semantic information successfully transmitted through the wireless channel. The second component, $P^0({\sum\limits_{i = 1}^C {{\overline w_i}} } - \sum\limits_{i = 1}^M {{\overline w_i}} )$, represents the inherent predictability of features that are not transmitted.

        Furthermore, to explore the relationship between $\eta $ and inference performance, we manipulate $M$ and $\rm{SNR}$, record the corresponding $\eta$ and the test inference performance, and subsequently apply curve fitting to derive the mapping function $\phi (\eta)$. The workflow is outlined in Algorithm \ref{alg:curve}.

        Note that the SNR is randomly sampled, and binary search \cite{knuth1997art} is employed to determine $M$, as $\eta$ monotonically increases with $M$ according to Eq.~\eqref{eq:STII}.
        Specifically, we perform a binary search within the interval $[1,C]$ and compute the STII according to Eq.~\eqref{eq:STII}. The search boundaries are iteratively adjusted based on the comparison between the computed and target STII values, which enables us to efficiently locate the minimum value of $M$ that approximates the target STII.
        
        \begin{algorithm}[h]
            \caption{Workflow of curve fitting for the mapping function $\phi(\eta)$}
            \label{alg:curve}
            \KwIn{Dataset $\left\{ {\left( {{{\bf{s}}_j},{{\bf{y}}_j}} \right)} \right\}_{j = 1}^N$ , Target STII $ \{ \eta_i \} _ {i=1} ^ S$}
            \KwOut{Mapping function $\phi(\eta)$}
            
            \For{$\eta_i$ in $\{ \eta_i \} _ {i=1} ^ S$}{
                \For{ ${\left( {{{\bf{s}}_j},{{\bf{y}}_j}} \right)}$ in $\left\{ {\left( {{{\bf{s}}_j},{{\bf{y}}_j}} \right)} \right\}_{j = 1}^N$}{
                    Sample $\rm{SNR}$ which meets $(1 - P_e(\rm{SNR})) < \eta$\;
                    Extract $\mathbf{x}$ from $\mathbf{s}$\;
                    Compute $w_k$ using Eq.~\eqref{eq:importance} and sort to obtain $\mathbf{\overline{w}}$\;
                    Apply binary search to determine $M$\;
                    Compute $\overline{\mathbf{x}}$ according to Eq.~\eqref{eq:filter}\;
                    Decode to get $\mathbf{\hat{y}}$ using Eq.~\eqref{eq:decode}\;
                    Store current inference performance and $\eta_i$\;
                }
                Compute average inference performance and $\eta_i$\;
            }
            Curve fitting for $\phi(\eta)$ using stored data and $\{ \eta_i \} _ {i=1} ^ S$\;
            
        \end{algorithm}



        Subsequently, the feature selection problem can be represented as finding the minimum $M$ in order to minimize the transmission rate while satisfying performance requirements:
        \begin{subequations}
            \begin{align}
            \mathop {\min }\limits_M & \frac{ML\log_2 U}{T_s} \\
            \mathrm{s.t.} &\eta \ge \eta^* \label{eq:constraint}\\ 
            &M \in \mathbb{Z}^+ \land M \leq  C, 
            \end{align}
            \label{eq:optimization}
        \end{subequations}

        \vspace{-5mm}
        where U denotes the modulation order, $T_s$ represents the average transmission interval between signals, and $\eta^*$ is the STII threshold obtained through the inverse function $\phi^{-1}(\tau)$, with $\tau$ denoting the inference performance threshold. This optimization problem can be efficiently solved using binary search. Furthermore, when the current BER is too high to meet the STII threshold requirement, $M$ will equal $C$ to maximize robustness of the system. The computational complexity of this module is $O(\log C)$.
        
        Furthermore, the workflow of the IRCSC scheme is summarized in Algorithm~\ref{alg:workflow}. 
        
        \begin{algorithm}[h]
            \caption{Workflow of the IRCSC scheme}
            \label{alg:workflow}
            \SetAlgoLined
            \KwIn{Input image $\mathbf{s}$, SNR, performance threshold $\tau$}
            \KwOut{Number of selected feature channels $M$, filtered features $\overline{\mathbf{x}}$}
        
            Extract and quantize semantic features $\mathbf{x}$ from input image  $\mathbf{s}$\;
        
            \For{each feature channel $k$}{
                Calculate importance weight $w_k$ based on Eq.~\eqref{eq:importance}\;
            }
            
            Calculate the target STII threshold $\eta^*$ from $\tau$\;
            $M \leftarrow$ Binary search for minimum $M$ based on \eqref{eq:optimization}\;
        
            Select feature channels to form $\overline{\mathbf{x}}$ using \eqref{eq:filter}\;
        
            \KwRet{$M$, $\overline{\mathbf{x}}$}
            \end{algorithm}

\section{Experimental Results and Analysis}
    In this section, we compare the proposed scheme with three baseline schemes to evaluate its effectiveness:

    \begin{itemize}
        \item \textbf{Traditional Deep JSCC (TD-JSCC) \cite{xie2021deep}}: This scheme does not implement rate adaptation, specifically, it transmits complete semantic features under any channel environment.

        \item \textbf{Without importance analyzer (WO-IA)}: This scheme shares the number of feature channels $M$ as the IRCSC method, but implements feature selection randomly without the assistance of $\overline{\mathbf{w}}$.

        \item \textbf{Without feature selector (WO-FS)}: This scheme follows a selection approach analogous to IRCSC using Eq.~\eqref{eq:filter}, but operates without the proposed feature selector module to compute the optimal $M$ based on channel conditions. Instead, it employs a fixed selection of $M=48$ feature channels regardless of the channel condition
        \item \textbf{Soft actor-critic based rate control semantic communication (SAC-RCSC)}: This scheme shares the importance analyzer with IRCSC, but determines $M$ through an RL policy $\pi_\theta$. The state space $\mathcal{S} = \text{SNR}$, and the action space is $\mathcal{A} = \{M \in \mathbb{Z}^+ | 1 \leq M \leq C\}$. The policy is optimized using the SAC algorithm.

    \end{itemize}

    We conduct experiments on the ImageNet \cite{ImageNet} dataset, which contains over 1.2 million training images and 50,000 validation images across 1,000 object categories. The semantic encoder and decoder employ a pre-trained swin transformer, which is jointly fine-tuned with the quantizer under a 0dB Rayleigh channel using Adam optimizer. Both the baselines and the IRCSC scheme share the same parameters. 
    
\begin{figure}[!h]
        \centering
        \includegraphics[width=0.72\linewidth]{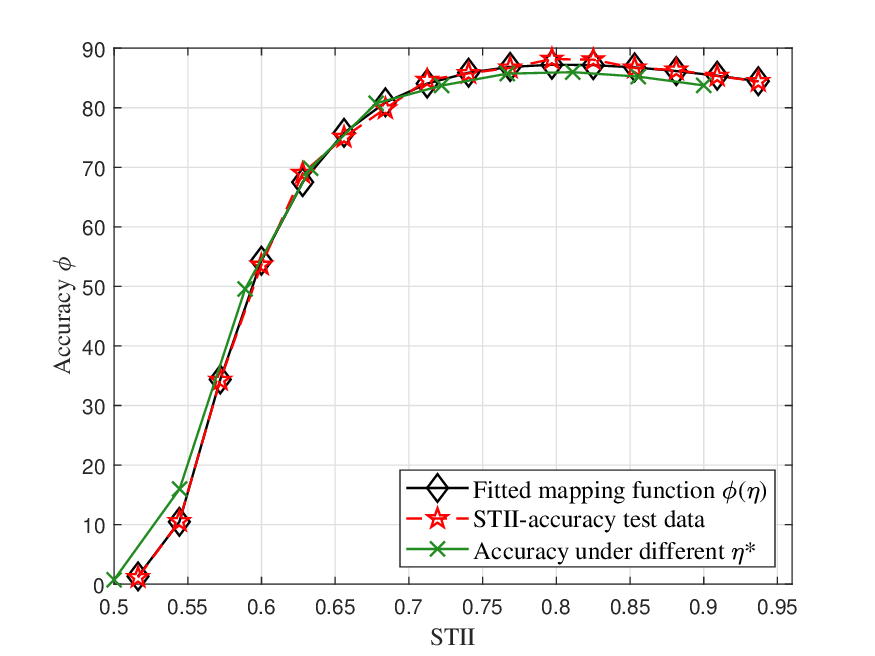}
        \caption{Comparison between the mapping function and the test data curves}
        \label{fig:curve}
    \end{figure}

    Based on the above settings, we generate the data required for fitting the mapping function $\phi(\eta)$ according to Algorithm~\ref{alg:curve}, where the channel environment is a Rayleigh fading channel with SNR varying from -10dB to 10dB. The fitting results are shown in Fig.~\ref{fig:curve}, where a rational function is selected as the fitting function:

    \begin{equation}
        \phi(\eta) =\frac{p_{1} \eta^{3}+p_{2} \eta^{2}+p_{3} \eta+p_{4}} {\eta^{2}+q_{1} \eta+q_{2}},
    \end{equation}
    where $ \{ p_i \} _ {i=1} ^ 4$ and $ \{ q_i \} _ {i=1} ^ 2$ are parameters of the mapping function, as listed in Table ~\ref{table:Parameter}. Moreover, by adjusting $\eta^*$ in \eqref{eq:constraint}, different performance-rate trade-off can be achieved. As shown in Fig. 2, setting $\tau$ to approximately 86 (corresponding to $\eta \approx 0.76$) achieves the optimal performance-rate trade-off. Furthermore, when channel resources are extremely constrained, controlled lower rate can still be achieved.

    \begin{table}[h]
        \caption{Parameters of mapping function $\phi(\eta)$}
        \centering
        \scalebox{0.8}{
            \begin{tabular}{|c|c|c|c|c|c|c|}
                \hline
                \rule{0pt}{15pt}Parameters & ${p_1}$ &${{p_2}}$ & ${{p_3}}$ & ${{p_4}}$&$ q_1 $ & $ q_2 $ \\ \hline
                \rule{0pt}{15pt}  Values
                &-60.34&210.9 & -170.9&40.3 & -1.021 &0.2652\\ \hline
            \end{tabular}
            \label{table:Parameter}
            }
        \end{table}
        
    Table \ref{tab:latency} presents the computational latency of different schemes tested on one NVIDIA RTX 4070Ti GPU. Compared to TD-JSCC, IRCSC requires an additional 0.31ms latency, with 0.15ms from the extra neural network computation of the feature extractor and 0.08ms from solving Eq.~\eqref{eq:optimization}. Compared to the SCAN-RCSC scheme, IRCSC achieves only one-sixth of its computational latency.

    \begin{table}[ht]
        \caption{Comparison of computational latency (ms)}
        \centering
        \begin{tabular}{|p{3cm}|c|c|c|}
            \hline
            \textbf{Component} & \textbf{TD-JSCC} & \textbf{SAC-RCSC} & \textbf{IRCSC} \\ \hline
            Neural Network Processing & 11.57 & 11.82  & 11.82 \\ \hline
            Algorithm Execution & N/A & 0.48 & 0.08 \\ \hline 
            Other Operations & 0.26 & 0.34 & 0.34 \\ \hline
            Total Latency & 12.08 & 12.54 & 12.24 \\ \hline
        \end{tabular}
        \label{tab:latency}
    \end{table}
 
\begin{figure}[h]
        \centering
        \begin{subfigure}[Accuracy over AWGN channel]{
            \includegraphics[width=0.72\linewidth]{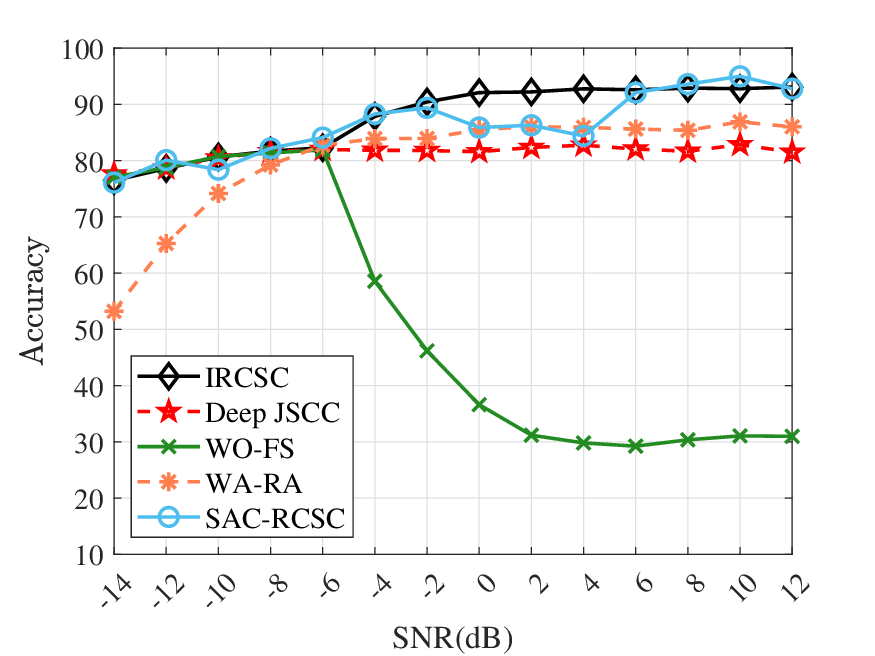}
            \label{fig:acc_awgn}}
        \end{subfigure}
        \begin{subfigure}[Accuracy over Rayleigh channel]{
            \includegraphics[width=0.72\linewidth]{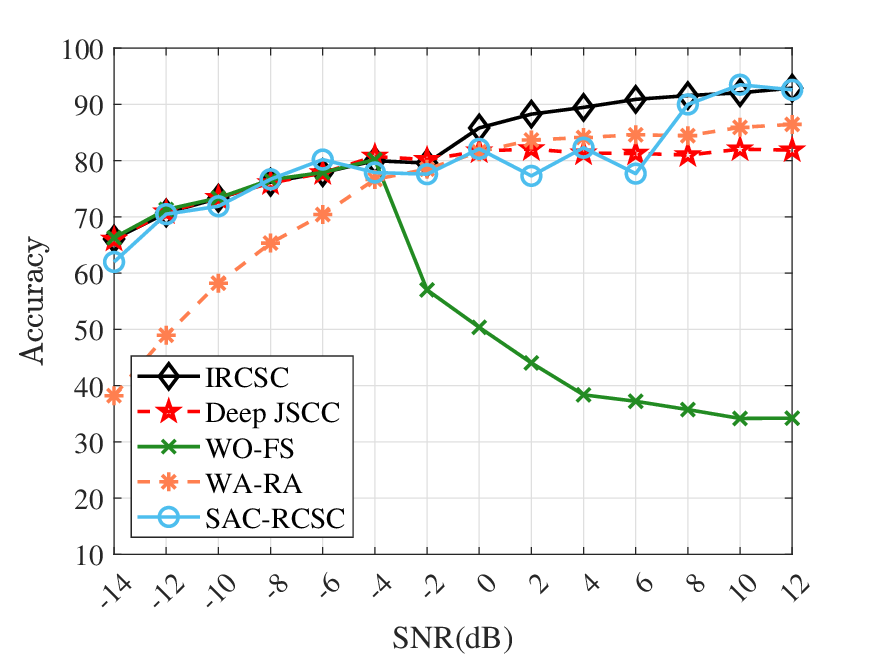}
            \label{fig:acc}}
        \end{subfigure}
        \caption{Accuracy comparison of different schemes over AWGN and Rayleigh channels}
    \end{figure}

    Fig.~\ref{fig:acc_awgn} and Fig.~\ref{fig:acc} show the performance comparison between the proposed scheme and three baselines over AWGN and Rayleigh channels, respectively. In the low SNR region, due to severe channel interference, the constraint requirements cannot be met, thus, the IRCSC scheme chooses to transmit the whole semantic feature, and achieving the same performance as TD-JSCC. As SNR increases, the IRCSC scheme begins transmission rate control and gradually outperforms TD-JSCC. Moreover, compared to the WO-IA scheme, our proposed method maintains performance while reducing the transmission rate. Compared to the scheme WO-FS, our approach avoids performance degradation in low SNR scenarios and maintains superior performance throughout the simulation. Compared with SAC-RCSC, IRCSC demonstrates more stable rate control performance.

\begin{figure}[h]
        \centering
        \begin{subfigure}[Transmission rate over AWGN channel]{
            \includegraphics[width=0.72\linewidth]{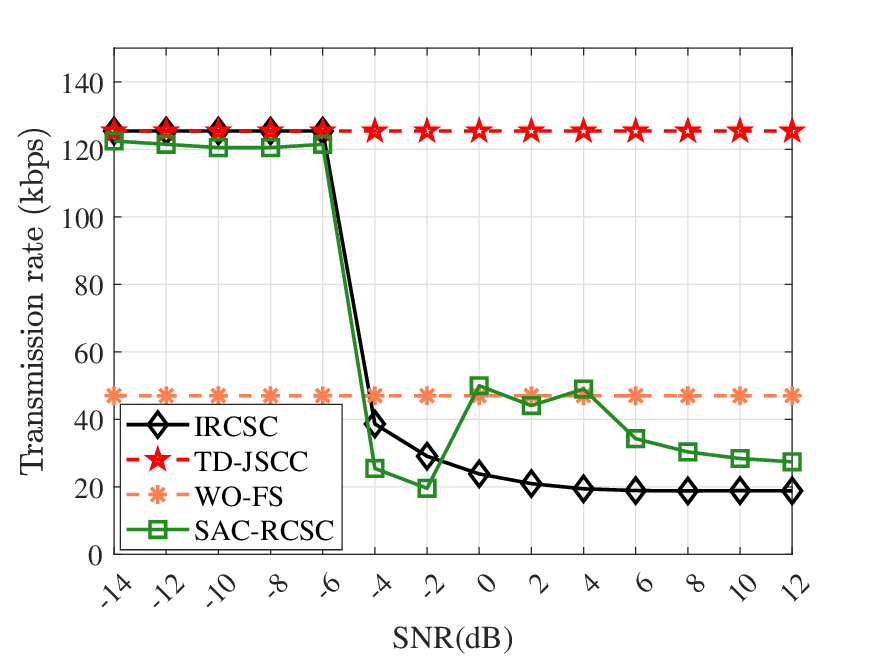}
            \label{fig:rate_awgn}}
        \end{subfigure}
        \begin{subfigure}[Transmission rate over Rayleigh channel]{
            \includegraphics[width=0.72\linewidth]{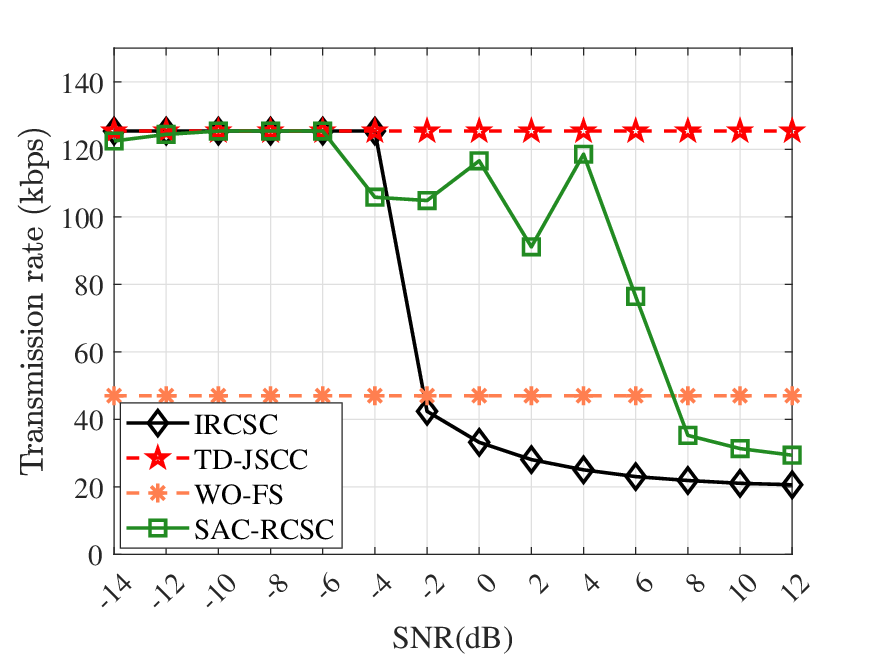}
            \label{fig:rate}}
        \end{subfigure}
        \caption{Transmission rate comparison of different schemes over AWGN and Rayleigh channels}
    \end{figure}

    Fig.~\ref{fig:rate_awgn} and Fig.~\ref{fig:rate} presents a comparative analysis of the transmission rates across schemes, with $T_s$ in Eq.~\eqref{eq:optimization} set to 50ms. The TD-JSCC scheme transmits at 125.44 kilobits per second (kbps), while the WO-FS scheme maintains a constant transmission rate of 47.04 kbps. For the AWGN channel, IRCSC scheme transmits the complete semantic features when the SNR is below -6dB, as the constraint Eq.~\eqref{eq:constraint} cannot be met. As the SNR exceeds -6dB, the transmission rate gradually decreases, reaching an average of 14.48 kbps at 12dB. For the Rayleigh channel, a similar trend can be observed with the transition point at -2dB, where the IRCSC scheme begins to reduce the transmission rate, reaching approximately 20.61 kbps at 12dB. At an SNR of 12 dB for both channels, the IRCSC scheme achieves a transmission rate less than one-sixth of that of the TD-JSCC scheme. Moreover, IRCSC has lower rates compared to SAC-RCSC in most channel environments, especially in Rayleigh channels.

\section{Conclusions}
We propose a novel IRCSC scheme for task-oriented semantic communication systems that dynamically adjusts transmission rates using an importance analyzer and feature selector based on channel conditions and semantic relevance. Experiments show our approach outperforms traditional Deep JSCC methods in high SNR scenarios with better performance and compression. Compared to conventional rate control techniques, our method offers lower computational complexity and flexible performance-rate trade-off management.

\bibliographystyle{IEEEtran}

\bibliography{reference}

\begin{thebibliography}{10}
\providecommand{\url}[1]{#1}
\csname url@samestyle\endcsname
\providecommand{\newblock}{\relax}
\providecommand{\bibinfo}[2]{#2}
\providecommand{\BIBentrySTDinterwordspacing}{\spaceskip=0pt\relax}
\providecommand{\BIBentryALTinterwordstretchfactor}{4}
\providecommand{\BIBentryALTinterwordspacing}{\spaceskip=\fontdimen2\font plus
\BIBentryALTinterwordstretchfactor\fontdimen3\font minus
  \fontdimen4\font\relax}
\providecommand{\BIBforeignlanguage}[2]{{%
\expandafter\ifx\csname l@#1\endcsname\relax
\typeout{** WARNING: IEEEtran.bst: No hyphenation pattern has been}%
\typeout{** loaded for the language `#1'. Using the pattern for}%
\typeout{** the default language instead.}%
\else
\language=\csname l@#1\endcsname
\fi
#2}}
\providecommand{\BIBdecl}{\relax}
\BIBdecl

\bibitem{akyildiz2022metaverse}
I.~F. Akyildiz, ``Metaverse: Challenges for extended reality and
  holographic-type communication in the next decade,'' in \emph{2022 ITU
  Kaleidoscope-Extended reality--How to boost quality of experience and
  interoperability}.\hskip 1em plus 0.5em minus 0.4em\relax IEEE, 2022, pp.
  1--2.

\bibitem{strinati20196g}
E.~C. Strinati, S.~Barbarossa, J.~L. Gonzalez-Jimenez, D.~Ktenas, N.~Cassiau,
  L.~Maret, and C.~Dehos, ``6g: The next frontier: From holographic messaging
  to artificial intelligence using subterahertz and visible light
  communication,'' \emph{IEEE Vehicular Technology Magazine}, vol.~14, no.~3,
  pp. 42--50, 2019.

\bibitem{dong2022semantic}
C.~Dong, H.~Liang, X.~Xu, S.~Han, B.~Wang, and P.~Zhang, ``Semantic
  communication system based on semantic slice models propagation,'' \emph{IEEE
  Journal on Selected Areas in Communications}, vol.~41, no.~1, pp. 202--213,
  2022.

\bibitem{shao2021learning}
J.~Shao, Y.~Mao, and J.~Zhang, ``Learning task-oriented communication for edge
  inference: An information bottleneck approach,'' \emph{IEEE Journal on
  Selected Areas in Communications}, vol.~40, no.~1, pp. 197--211, 2021.

\bibitem{liu2024explainable}
C.~Liu, C.~Guo, Y.~Yang, W.~Ni, Y.~Zhou, L.~Li, and T.~Q. Quek, ``Explainable
  semantic communication for text tasks,'' \emph{IEEE Internet of Things
  Journal}, 2024.

\bibitem{jankowski2020deep}
M.~Jankowski, D.~G{\"u}nd{\"u}z, and K.~Mikolajczyk, ``Deep joint
  source-channel coding for wireless image retrieval,'' in \emph{ICASSP
  2020-2020 IEEE International Conference on Acoustics, Speech and Signal
  Processing (ICASSP)}.\hskip 1em plus 0.5em minus 0.4em\relax IEEE, 2020, pp.
  5070--5074.

\bibitem{weng2023deep}
Z.~Weng, Z.~Qin, X.~Tao, C.~Pan, G.~Liu, and G.~Y. Li, ``Deep learning enabled
  semantic communications with speech recognition and synthesis,'' \emph{IEEE
  Transactions on Wireless Communications}, vol.~22, no.~9, pp. 6227--6240,
  2023.

\bibitem{li2023fast}
P.~Li, G.~Cheng, J.~Kang, R.~Yu, L.~Qian, Y.~Wu, and D.~Niyato, ``Fast:
  Fidelity-adjustable semantic transmission over heterogeneous wireless
  networks,'' in \emph{ICC 2023-IEEE International Conference on
  Communications}.\hskip 1em plus 0.5em minus 0.4em\relax IEEE, 2023, pp.
  4689--4694.

\bibitem{liu2024ofdm}
C.~Liu, C.~Guo, Y.~Yang, W.~Ni, and T.~Q. Quek, ``Ofdm-based digital semantic
  communication with importance awareness,'' \emph{IEEE Transactions on
  Communications}, 2024.

\bibitem{wang2024feature}
Y.~Wang, S.~Han, X.~Xu, H.~Liang, R.~Meng, C.~Dong, and P.~Zhang, ``Feature
  importance-aware task-oriented semantic transmission and optimization,''
  \emph{IEEE Transactions on Cognitive Communications and Networking}, 2024.

\bibitem{wang2025channel}
B.~Wang, R.~Gu, W.~Xu, F.~Jiang, M.~Li, and S.~Wang, ``Channel-aware deep joint
  source-channel coding for multi-task oriented semantic communication,''
  \emph{IEEE Wireless Communications Letters}, 2025.

\bibitem{selvaraju2017grad}
R.~R. Selvaraju, M.~Cogswell, A.~Das, R.~Vedantam, D.~Parikh, and D.~Batra,
  ``Grad-cam: Visual explanations from deep networks via gradient-based
  localization,'' in \emph{Proceedings of the IEEE international conference on
  computer vision}, 2017, pp. 618--626.

\bibitem{knuth1997art}
D.~E. Knuth, \emph{The Art of Computer Programming, volume 1 Fundamental
  Algorithms}.\hskip 1em plus 0.5em minus 0.4em\relax Addison Wesley Longman
  Publishing Co., Inc., 1997.

\bibitem{xie2021deep}
H.~Xie, Z.~Qin, G.~Y. Li, and B.-H. Juang, ``Deep learning enabled semantic
  communication systems,'' \emph{IEEE Transactions on Signal Processing},
  vol.~69, pp. 2663--2675, 2021.

\bibitem{ImageNet}
J.~Deng, W.~Dong, R.~Socher, L.-J. Li, K.~Li, and L.~Fei-Fei, ``Imagenet: A
  large-scale hierarchical image database,'' in \emph{2009 IEEE Conference on
  Computer Vision and Pattern Recognition}, 2009, pp. 248--255.

\end{thebibliography}

\end{document}